\documentclass[twocolumn,times,twocolappendix]{aastex631}

\usepackage{comment}

\begin{document}

\title{IMPACTX: An X-ray Spectral Model for Polar Dust and Clumpy Torus}

\author[0009-0006-4377-4219]{Kanta Fujiwara}
\affiliation{Department of Astronomy, Kyoto University, Sakyo-ku, Kyoto, Japan}

\author[0000-0002-5485-2722]{Yoshihiro Ueda}
\affiliation{Department of Astronomy, Kyoto University, Sakyo-ku, Kyoto, Japan}

\author[0000-0002-5701-0811]{Shoji Ogawa}
\affiliation{Institute of Space and Astronautical Science (ISAS), Japan Aerospace Exploration Agency (JAXA) 3-1-1 Yoshinodai, Chuo-ku, Sagamihara, Kanagawa, Japan}

\author[0009-0000-9577-8701]{Yuya Nakatani}
\affiliation{Department of Astronomy, Kyoto University, Sakyo-ku, Kyoto, Japan}

\author[0000-0001-6653-779X]{Ryosuke Uematsu}
\affiliation{Department of Astronomy, Kyoto University, Sakyo-ku, Kyoto, Japan}



\begin{abstract}

We construct a generic X-ray spectral model for the reflection
component from the clumpy torus and dusty gas in the polar region
(polar dusty gas) in an active galactic nucleus (AGN), designated
as Inclusive spectral energy distribution Model of
Polar dust And Clumpy Torus for X-ray (IMPACTX).
To calculate the spectra, we utilize the Monte-Carlo based, 3-dimensional
radiative transfer code SKIRT. The adopted geometry is the same as
that of the IMPACT model (Ogawa et al.\ in prep.), consisting of a
clumpy torus defined by Nenkova et al. (2008) and paraboloid-shell
polar dusty gas launched at the inner radius of the
torus. We discuss the dependence of the X-ray spectrum on geometrical
parameters in comparison with the conventional torus-only model
(XCLUMPY).  As an example, we have simultaneously applied the IMPACTX
and IMPACT models to the X-ray and infrared spectra of the nearby
Seyfert 2 galaxy NGC 4388, and have found a solution that can well
reproduce both spectra. This demonstrates the importance of using both
X-ray and infrared data to constrain the nuclear structure of an AGN.

\end{abstract}



\section{Introduction} 

In the present universe, the mass of the stellar bulge of a
galaxy and that of the supermassive black hole (SMBH) in its center
are tightly correlated, despite their large difference in spatial
scale by $\sim$10 orders of magnitude \citep{2013ARA&A..51..511K}.
The co-moving density of star formation rate (e.g., \citealt{2014ARA&A..52..415M})
and that of mass accretion rate onto SMBHs (e.g., \citealt{2014ApJ...786..104U, 2015MNRAS.451.1892A})
show overall similar cosmological evolution, peaking at
$z\sim2$. These facts lead to the idea of ``co-evolution'' of galaxies
and SMBHs. Active galactic nuclei (AGNs) are the key object to
investigate the physical processes of galaxy-SMBH co-evolution,
because it is the site where mass fed by the host galaxy is accreted
by the SMBH and also, because they are bright, AGNs allow you to actually observe SMBHs. To understand the mass transfer processes, it is critical to constrain the nuclear structure of AGNs.

In the classical unified model (\citealt{1985ApJ...297..621A, 1991PASJ...43..195A}) of
AGNs, the SMBH and accretion disk are surrounded by ``dusty torus''
(see \citet{2017NatAs...1..679R} for a review), which blocks the observer's
line-of-sight toward the SMBH when viewed edge-on. This model is
successful to explain many basic aspects of AGNs, particularly the
observed differences between type-1 (unobscured) and type-2 (obscured)
AGNs.  Recent observational and theoretical studies have revealed that
the nuclear structure is far more complex, however. Mid-infrared (IR)
interferometry observations of nearby AGN revealed that a significant
part of the dust emission extends in the polar direction, instead of
the equatorial direction where the classical torus resides
(e.g. \citealt{2004Natur.429...47J}; \citealt{2006A&A...452..459H,2012ApJ...755..149H,2013ApJ...771...87H};
\citealt{2014A&A...563A..82T};
\citealt{2014A&A...565A..71L,2016A&A...591A..47L};
\citealt{2018ApJ...862...17L}; \citealt{2019MNRAS.489.2177A}; \citealt{2022A&A...663A..35I,2025arXiv250201840I}; see
\citet{2016SPIE.9907E..0RB} for a review).
\ifnum0=1
More recently, the dust component of NGC 1068 in the polar direction
was imaged in the near infrared band by the Large Binocular
Telescope Interferometer (LBTI) \citep{2025arXiv250201840I}.
The Atacama Large Millimeter/submillimeter Array
(ALMA) also reveals polar outflows on the same spatial
scales in NGC 1068 (e.g. \citealt{2016ApJ...829L...7G}).
\fi
This structure is often called ``polar dust'' (in this paper we refer
to it as ``polar dusty gas'' because the total mass is dominated by
gas accompanying dust that emits mid-IR radiation). In fact, theories
predict the presence of a radiation-driven outflow launched at the
innermost region (\citealt{2014MNRAS.445.3878S,2016ApJ...828L..19W}), which most likely corresponds to the observed polar dusty gas.
On the basis of these results, \citet{2021ApJ...906...84O}
propose an updated unified scheme where the presence of polar dusty
gas is universal, to consistently explain X-ray and infrared
properties of AGNs. 

Infrared and X-ray spectra provide us with complementary information
on the torus structure including polar dusty gas. Infrared emission
directly traces AGN-heated dust, whereas X-rays trace all matter
including gas and dust through absorption and/or reflection signals.
It is important to have theoretical spectral models for AGNs,
which allow one to directly compare with the observations. Before the
discovery of polar dusty gas, models took into account only the
classical torus (i.e., without polar dusty gas) both in the infrared band
(e.g., \citealt{1988ApJ...329..702K}; \citealt{1992ApJ...401...99P}; \citealt{1995MNRAS.277.1134E}; \citealt{2005A&A...436L...5S}; \citealt{2006MNRAS.366..767F}; \citealt{2008ApJ...685..147N, 2008ApJ...685..160N}; \citealt{2008A&A...482...67S}; \citealt{2010A&A...523A..27H};  \citealt{2012MNRAS.420.2756S,2016MNRAS.458.2288S}) and in the X-ray band (e.g., \citealt{2009MNRAS.397.1549M,
2009ApJ...692..608I, 2017A&A...607A..31P, 2018ApJ...854...42B} for smooth tori; e.g.,
\citealt{2016ApJ...818..164F, 2019ApJ...877...95T}, and \citealt{2019A&A...629A..16B} for clumpy tori.
Given the fact that the polar dusty gas is ubiquitous in AGNs, these models
must be updated by including the contribution from it.

\citet{2023ApJ...945...55R} also point out the importance of considering dusty media in high-resolution X-ray spectroscopy, which produces complex near-edge X-ray absorption fine structures.

In the infrared band, \citet{2017ApJ...838L..20H} calculated infrared spectra from
the torus and polar dusty gas using a radiative transfer model called
CAT3D-WIND. \citet{2017MNRAS.472.3854S,2019MNRAS.484.3334S} studied the infrared spectrum from polar
dusty gas in Circinus Galaxy, using the SKIRT code (\citealt{2011ascl.soft09003B}; \citealt{2015A&C.....9...20C}; \citealt{2020A&C....3100381C}; \citealt{2023A&A...674A.123V}). Recently,
Ogawa et al.\ (in preparation. hereafter Paper~I) have constructed a
more generic infrared spectral model using SKIRT. They are able to
well reproduce the infrared spectra of 28 nearby AGNs whose nuclei are
separated by ground-based, high angular-resolution mid-infrared
imaging (\citealt{2015ApJ...803...57I, 2019MNRAS.486.4917G}). This model, designated as Inclusive
spectral energy distribution Model of Polar dust And Clumpy Torus
(IMPACT), adopt the same geometry as in the CLUMPY
(\citealt{2008ApJ...685..147N, 2008ApJ...685..160N}) and XCLUMPY
\citep{2019ApJ...877...95T} models for the torus, and the
paraboloid-shell geometry for the polar dusty gas, which is similar to
hourglass-like curved shell suggested by \citet{2016ApJ...828L..19W}
and \citet{2023ApJ...950...72K}.

In the X-ray bands, several authors studied the reflection component
from the polar dusty gas. \citet{2019MNRAS.490.4344L} simulated the
reflection spectra from the clumpy torus and clumpy polar dusty gas,
and suggested that fluorescent lines below 3 keV may significantly
come from the polar dusty gas.  \citet{2022MNRAS.512.2961M} also
investigated the X-ray spectral contribution of polar dust with
different geometries. They also reported that optically thin polar
dust increases the scattering component at the low energy side of the
X-ray spectrum.
\citet{2022MNRAS.511.5768A} presented an X-ray spectral model
specifically for Circinus Galaxy, assuming a more complex geometry, including the accretion disk, flared disks, broad line region, torus, and polar dusty gas.  
These works, though novel, assumed very 
simplified geometry for the polar dusty gas (hollow cone or filled cone), which may not be compatible with the infrared SED (\citealt{2019MNRAS.484.3334S}, Paper~I).
In addition, the calculations were performed only within a limited parameter range and hence cannot be applied to many objects.

In this paper, we construct a more generic X-ray spectral model from
the clumpy torus and polar dusty gas that is applicable to any
objects, designated as Inclusive SED Model of Polar dust and Clumpy
Torus for X-ray (IMPACTX). We adopt the same geometry as in the IMPACT
model (Paper~I). This enables us to simultaneously analyze the X-ray
and infrared spectra under a common assumption for the geometry.  The
structure of this paper is as follows. Section
\ref{section:simulation} describes the adopted geometry of the torus
and polar dusty gas, and the details of Monte Carlo simulations. In
section \ref{section:results}, we report the basic properties of our
model, and discuss the effects of including the polar dusty gas with
various geometrical parameters by comparison with the XCLUMPY
model.
In section \ref{section:NGC4388}, we simultaneously analyze the
X-ray spectra and infrared SED of the nearby Seyfert-2 galaxy NGC 4388,
using IMPACTX and IMPACT, respectively. 
Throughout the paper, errors correspond to the 90$\%$
confidence region for a single parameter.

\section{Simulations} \label{section:simulation}

\subsection{Torus Geometry} 

In our model, we assume a clumpy torus, which is considered to be more
realistic case than a smooth torus with a uniform density (for a
brief review on this issue, see e.g., \citealt{2019ApJ...877...95T}).
The torus geometry is the same as in the
CLUMPY (\citealt{2008ApJ...685..147N, 2008ApJ...685..160N}) and XCLUMPY
\citep{2019ApJ...877...95T} models. It consists of randomly placed
clumps with a fixed size (the radius of each clump; $R_{\mathrm{clump}}$ is 0.002 pc)
according to a power-law distribution in the radial direction
and a Gaussian distribution in the angular direction. The number
density function $d(r, \theta, \phi)$ is represented in the spherical coordinate system (where $r$ is radius, $\theta$ is polar angle, and $\phi$ is azimuth) as: 
\begin{equation} 
  d(r, \theta, \phi) = N{\biggl ( \frac{r}{r_{\mathrm{in}}} \biggr ) }^{-q} \exp{\biggl (- \frac{{(\theta - \pi/2)}^2}{\sigma^2} \biggr ).}
\end{equation}
where N is the normalization, $r_{\mathrm{in}}$ is the inner radius of the torus, q is the index of the radial density profile, and $\sigma$ represents the torus angular width. 

The normalization N is related to the number of clumps along the equatorial plane 
$N^{\mathrm{Equ}}_{\mathrm{clump}}$ as

\begin{equation}
N^{\mathrm{Equ}}_{\mathrm{clump}} = \int^{r_{\mathrm{out}}}_{r_{\mathrm{in}}}d \biggl(r, \frac{\pi}{2}, 0 \biggr)\pi {R_{\mathrm{clump}}^2} dr,
\end{equation}

\begin{equation}
N = \frac{(1 - q)N^{\mathrm{Equ}}_{\mathrm{clump}}}{\pi {R_{\mathrm{clump}}^2} {r_{\mathrm{in}}^q}(r_{\mathrm{out}}^{1-q} - r_{\mathrm{in}}^{1-q} )}.
\end{equation}
where $r_{\mathrm{in}}$ and $r_{\mathrm{out}}$ are inner and outer radii of the torus.

Furthermore, the hydrogen column density along the equatorial plane $({N^{\mathrm{Equ}}_{\mathrm{H}}})$ is expressed using hydrogen number density $n_{\mathrm{H}}$ by the following equation.

\begin{equation}
{N^{\mathrm{Equ}}_{\mathrm{H}}} = \frac{4}{3} R_{\mathrm{clump}}N^{\mathrm{Equ}}_{\mathrm{clump}}n_{\mathrm{H}}
\end{equation}

We fix $N^{\mathrm{Equ}}_{\mathrm{clump}}$ to 10, $q = 0.5$, $r_{\mathrm{in}} = 0.05 ~ \mathrm{pc}$ and $r_{\mathrm{out}} = 1.00 ~ \mathrm{pc}$ as in the XCLUMPY model \citep{2019ApJ...877...95T}. Thus, the free parameters are the angular width of the torus  $(\sigma)$ and the hydrogen column density along the equatorial plane $({N^{\mathrm{Equ}}_{\mathrm{H}}})$.


\subsection{Geometry of Polar Dusty Gas} 

\begin{figure}[htb]
\centering
\begin{minipage}[h]{1.0\columnwidth}
    \centering
    \includegraphics[width=0.95\columnwidth]{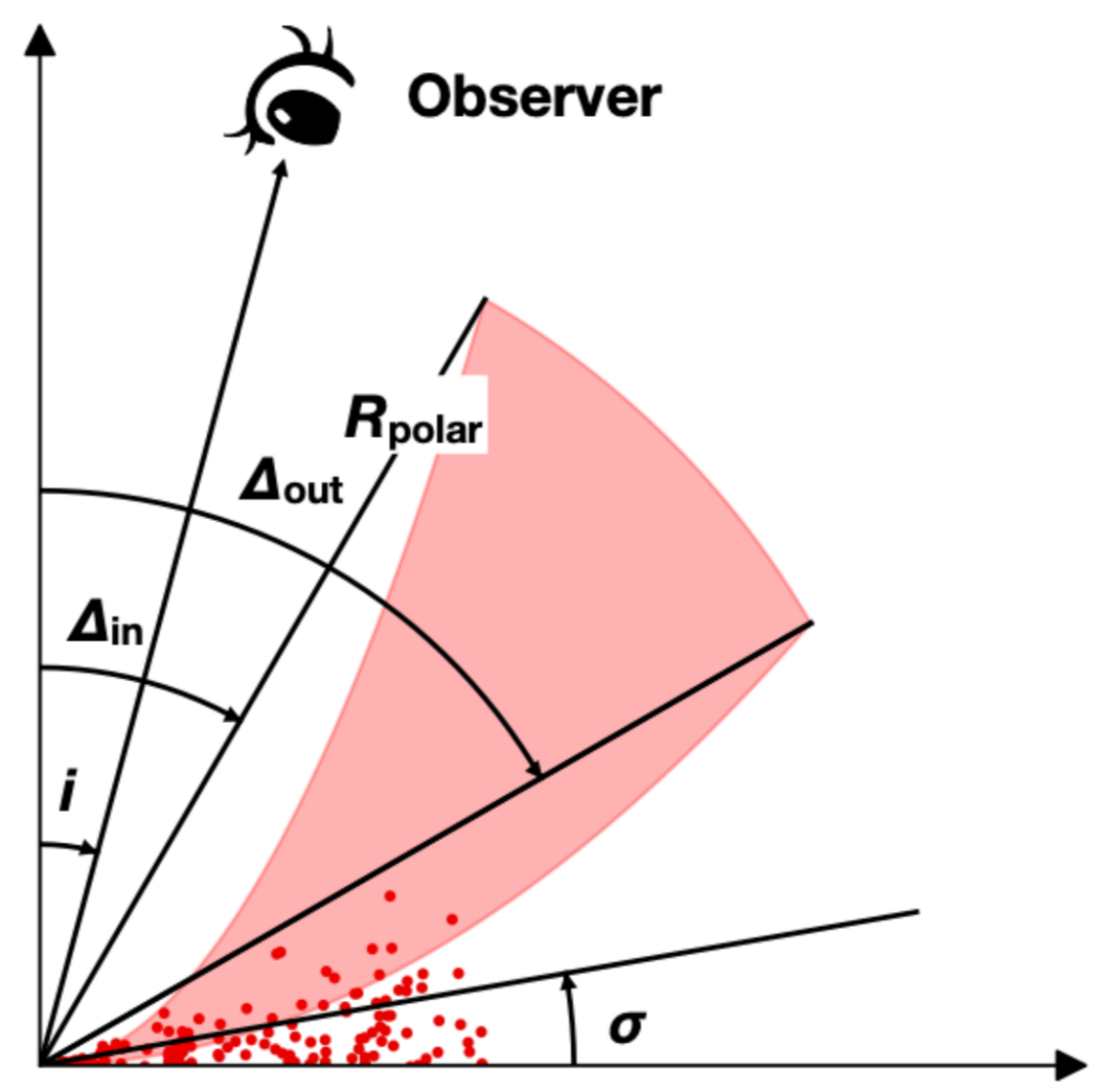}
\end{minipage}
    \caption{Cross-section view of the torus and polar dusty gas (taken from Figure 1 of Paper~I). }
    \label{figure:geometry}
\end{figure}

For the polar dusty gas, we adopt the same paraboloid-shell geometry
(Figure~\ref{figure:geometry}) as that in the IMPACT model
(Paper~I). The shape is similar to hourglass-like curved shell
suggested by \citet{2016ApJ...828L..19W} and
\citet{2023ApJ...950...72K} based on their three dimensional
radiation-hydrodynamic simulations.
It has 5 geometric parameters: radial extent of the paraboloid shell $R_{\mathrm{Polar}}$, half opening angle of the inner paraboloid $\Delta_{\mathrm{in}}$, that of the outer paraboloid $\Delta_{\mathrm{out}}$, and the inner radius of the paraboloid shell $r_{\mathrm{in}}$. In this work, we fix $\Delta_{\mathrm{out}}$ to $\Delta_{\mathrm{in}} + 30$ deg, which well reproduces typical IR SEDs of AGNs (Paper~I), 
and $r_{\mathrm{in}}$ to 0.05 pc, for consistency with the torus inner radius adopted in XCLUMPY.
%
For simplicity,  we assume 
uniform density distribution of gas and dust as in the IMPACT model.
  To make the spectral model ``scale-free'', i.e., invariant for any choice of $r_{\mathrm{in}}$, which depends on the bolometric luminosity of AGN (Paper~I),  we define the radial-thickness parameter of the paraboloid
  $Y_{\mathrm{Polar}} = R_{\mathrm{Polar}}/r_{\mathrm{in}}$ and the
  hydrogen column-density
parameter $N_\mathrm{H}^{\mathrm{Polar}} = R_{\mathrm{Polar}}\,
n_\mathrm{H}^{\mathrm{Polar}}$
\footnote{Hereafter we simply refer to $N_\mathrm{H}^{\mathrm{Polar}}$ as
  the hydrogen column density (of the polar dusty gas), which is an effective value integrated along a straight line from the origin to $R_{\mathrm{Polar}}$ assuming a constant density.}
where $n_\mathrm{H}^{\mathrm{Polar}}$
is the hydrogen number density.
Thus, the free parameters specific to the polar dusty gas are $\Delta_{\mathrm{in}}$,
$Y_{\mathrm{Polar}}$, and $N_\mathrm{H}^\mathrm{Polar}$.

In this study, we restrict
$N_\mathrm{H}^{\mathrm{Polar}} < 10^{23.5}~\mathrm{cm}^{-2}$; otherwise the photoelectric absorption by the polar dusty gas becomes unrealistically large even in type-1 AGNs, which is contradictory to the observations \citep{2019MNRAS.490.4344L}.
Observational studies have shown that the extent of polar dust ranges from 10 to several hundred parsecs (e.g. \citealt{2000AJ....120.2904B},
\citealt{2003ApJ...587..117R}, \citealt{2005ApJ...618L..17P},
\citealt{2016ApJ...822..109A}, \citealt{2025arXiv250201840I}).
Thus, we limit $Y_{\mathrm{Polar}} <2000$, corresponding to
$R_{\mathrm{Polar}} < 800 $ pc for $r_{\rm in} = 0.4$ pc, the dust sublimation radius for a typical bolometric luminosity of $10^{45}$ erg s$^{-1}$ (\citealt{2008ApJ...685..147N, 2008ApJ...685..160N}).

\subsection{Material Properties and Physical Processes}

To produce the X-ray spectral model reflected from the torus and polar
dusty gas, we utilize the Monte-Carlo 3-dimensional simulation code
SKIRT (\citealt{2011ascl.soft09003B}; 
\citealt{2015A&C.....9...20C}; \citealt{2020A&C....3100381C};  \citealt{2023A&A...674A.123V}; \citealt{2024A&A...688L..33V}), which
can calculate over a wide range of wavelengths from X-rays to radio
waves.  We assume that all matter in the torus and polar dusty gas is
neutral (not ionized) gas. 
For simplicity, the effect of dust on the X-ray
spectra, such as the X-ray absorption fine structure, is ignored\footnote{Nevertheless, we call
    this component as ``polar dusty gas'' throughout the paper, to distinguish it from ionized, dust-free gas
    in the polar region.
}. The
physical processes considered are photoelectric absorption, electron
scattering, and fluorescence emission lines. In SKIRT, the
photoelectric-absorption cross sections by \citet{1995A&AS..109..125V}
and \citet{1996ApJ...465..487V} are adopted, and scattering by
electrons bound in gas is taken into account. For our model
calculation, we assume the solar abundances by \citet{2009LanB...4B..712L}.

In our SKIRT simulations, we adopt a cubic spatial domain extending $4\times R_{\mathrm{Polar}}$ pc in all three directions, discretized using an octree-based adaptive grid 
with a maximum subdivision level of 15, corresponding to a minimum cell size of $\sim$0.0012 pc within the 40 pc simulation domain. This is finer than the diameter of the smallest structural elements in the model—the clumps (0.004 pc)—and is therefore adequate for resolving them. The spectral energy grid for the instruments is defined on a logarithmic scale, covering the range from 0.5 to 100 keV with 4000 grid points. The energy grid around 6.4 keV has a spacing of approximately 0.01 keV, which provides sufficient spectral resolution for comparison with typical X-ray CCD observations. For each run, $5\times10^6$
photon packets are launched. The intrinsic photon spectrum is modeled by the form of 
$A E^{-\Gamma} {\rm exp}(-E/E_{\rm cut})$, where $A$ is the normalization, $\Gamma$ the photon index, and $E_{\rm cut}$ the high energy cutoff. The simulations were performed over the parameter grid summarized in Table~\ref{c4grid}.

\begin{deluxetable*}{llll}[htb]
\tablecaption{Summary of Parameters\label{c4grid}}
\tablewidth{0pt}
\tablehead{No. &Parameter  &Grid & Units
}
\startdata
(1)&$\sigma$  &   10, 20, 30 & degree \\
(2)&$\log{N_\mathrm{H}^\mathrm{Equ}}$  &   23.5, 24.0, 24.5 & $\mathrm{cm}^{-2}$ \\
(3)&$\Delta_\mathrm{in}$ & 10, 20, 30, 40 & degree\\
(4)&$Y_{\mathrm{Polar}}$  & 
 200, 2000& \\
(5)&$\mathrm{\log{N_\mathrm{H}^\mathrm{Polar}}}$ &21.5, 22.0, 22.5, 23.0, 23.5&cm$^{-2}$\\
(6)&$Z(\mathrm{Fe})$ &  1.0, 1.5, 2.0, 2.5, 3.0 & solar\\
(7)&$i$  &  65.0, 75.0, 85.0 & degree\\
\enddata
\tablecomments{
(1) Half angular width of the torus.
(2) Logarithm of the hydrogen column density of the torus along the equatorial plane . 
(3) Half opening angle of inner wall of the polar dusty gas.
(4) Radial thickness parameter of the polar dusty gas ($R_{\mathrm{Polar}}/r_{\mathrm{in}}$).
(5) Logarithm of the hydrogen column density of the polar dusty gas.
(6) Abundance of Fe relative to hydrogen.
(7) Inclination angle of the observer.
}
\end{deluxetable*}

\section{Results}\label{section:results}

\begin{figure*}[htb]
\centering
\includegraphics[width=0.60\textwidth]{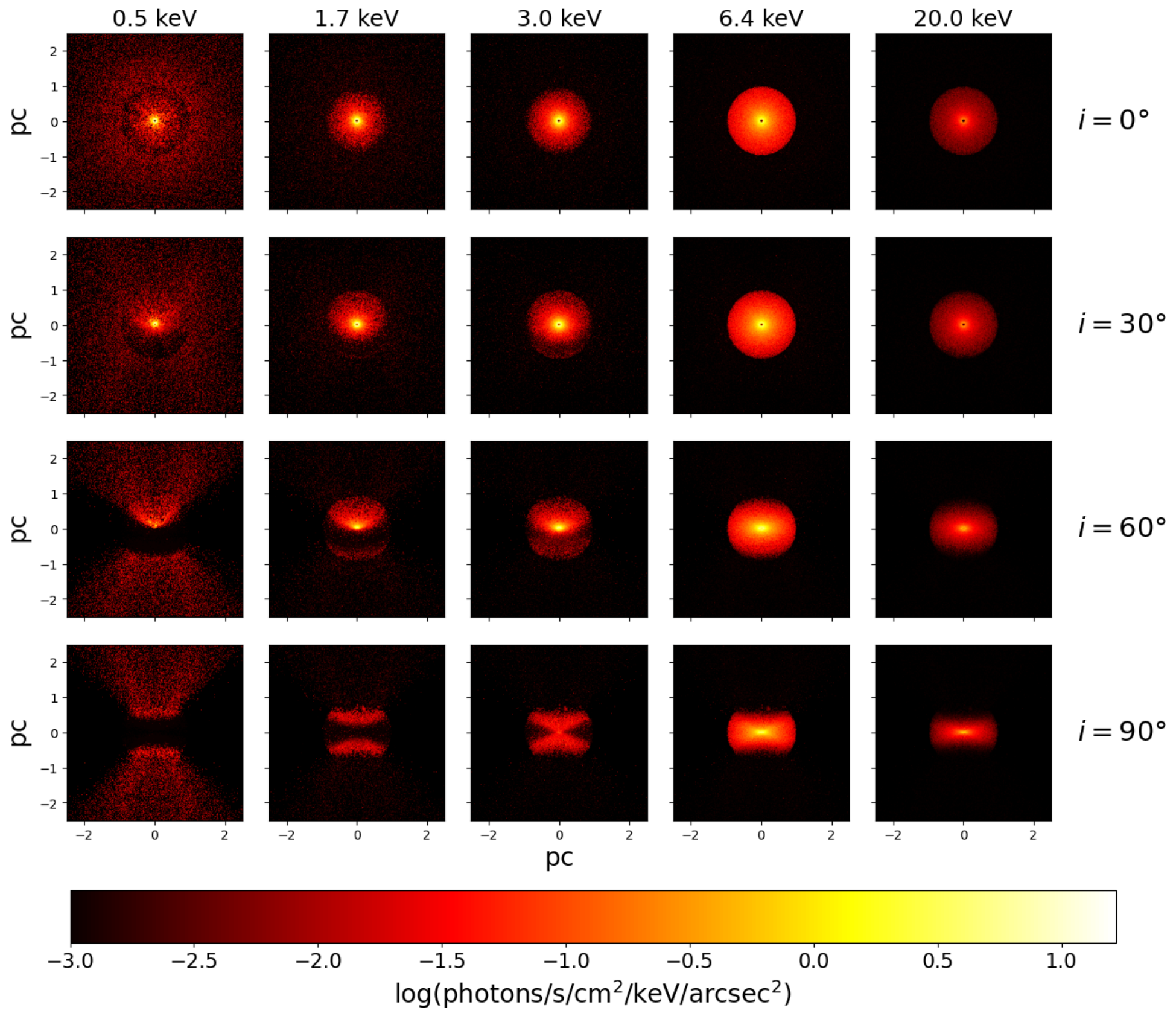}
\caption{Simulated images of the torus plus polar dusty gas around 0.5 $\mathrm{keV}$, 1.7 $\mathrm{keV}$, 3.0 $\mathrm{keV}$, 6.4 $\mathrm{keV}$, 20.0 $\mathrm{keV}$ viewed at inclination angles of i = $0^{\circ}$, $30^{\circ}$, $60^{\circ}$, and $90^{\circ}$ (from top to bottom).
The color bar represents the photon flux density per $\mathrm{arcsec}^2$.  We adopt the
following parameters: $\log{N_\mathrm{H}^{\mathrm{Torus}}}/{\mathrm{cm}}^{-2} = 24.0$,
$\sigma = 20.0^\circ$, $\Gamma = 1.8$, $E_{\mathrm{cut}} =
370 ~\mathrm{keV}$, ${\Delta}_\mathrm{{in}} = 10.0^\circ$,
$Y_{\mathrm{Polar}} = 200$, 
and 
$\log{N_\mathrm{H}^\mathrm{Polar}}/{\mathrm{cm}}^{-2} = 21.5$.
\label{figure:scattered}}
\end{figure*}

\begin{figure*}
\centering
\includegraphics[width=0.45\textwidth]{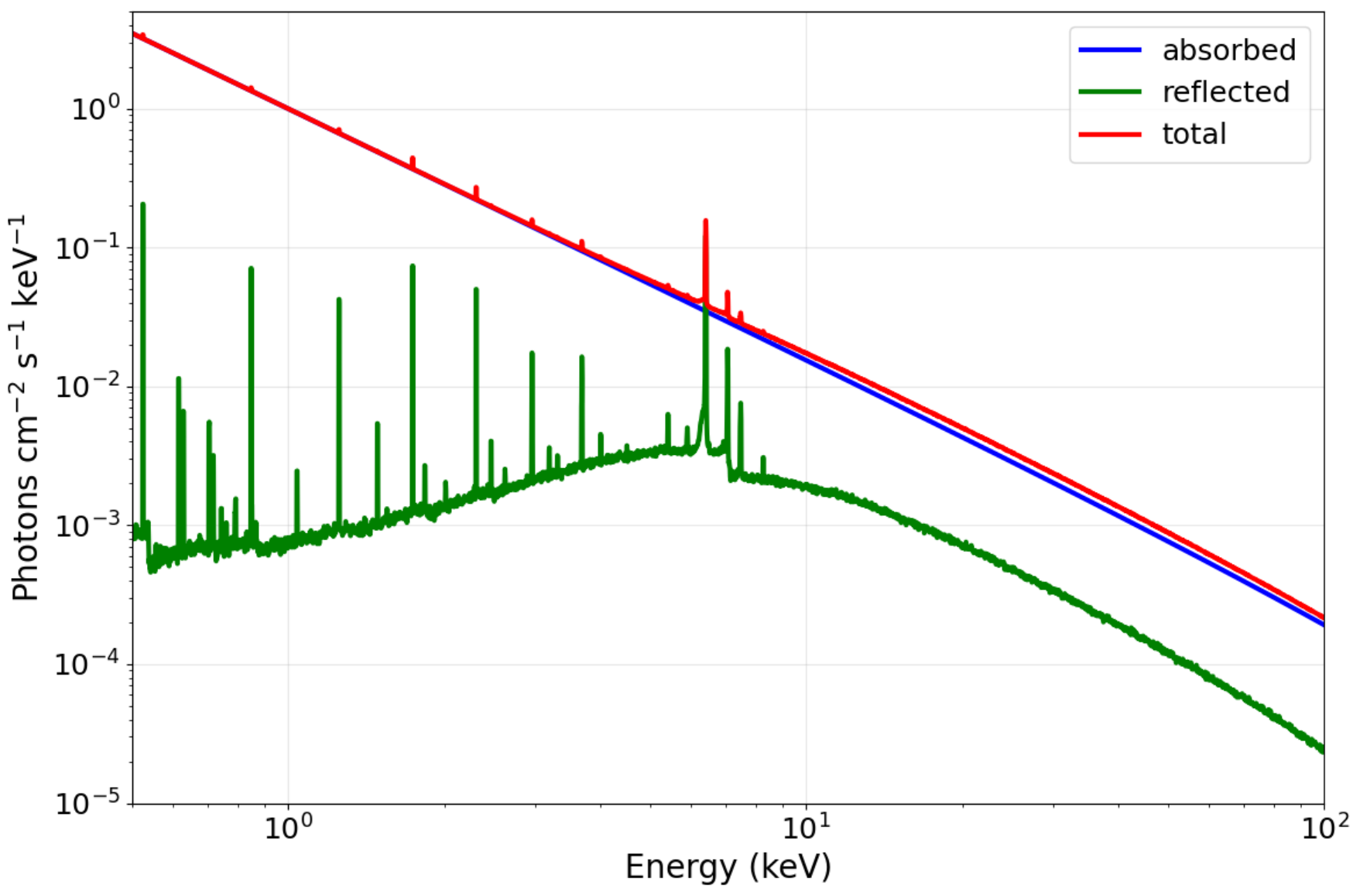}
\includegraphics[width=0.45\textwidth]{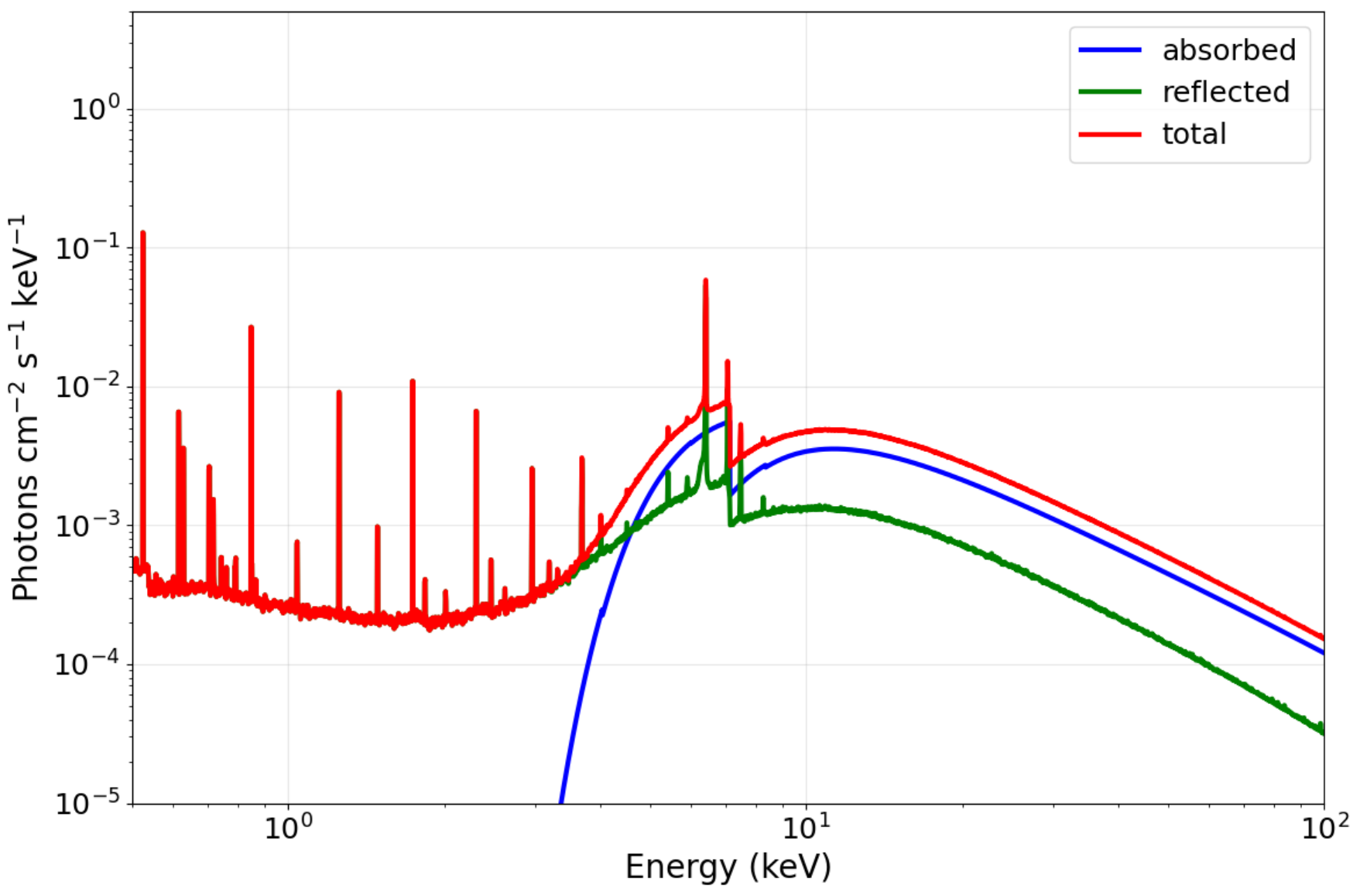}
\caption{
 Comparison of the reflection and direct components at $i = 10.0^\circ$ (left) and at $i = 80.0^\circ$ (right).
 We assume $\log{N_\mathrm{H}^{\mathrm{Torus}}}/{\mathrm{cm}}^{-2} = 24.0$, $\sigma = 20.0^\circ$, $\log{N_\mathrm{H}^\mathrm{Polar}}/{\mathrm{cm}}^{-2} = 21.5$, $\Delta_{\mathrm{in}} = 10.0^\circ$, $Y_{\mathrm{Polar}} = 200$, $\Gamma =
  1.8$, and $E_{\mathrm{cut}} = 370 ~\mathrm{keV}$.}
 \label{figure:model_i10}
 \label{figure:model_i80}
\end{figure*}

We investigate the dependence of the reflected X-ray spectrum on the
polar dusty gas parameters. Here we adopt the following default set of
parameters, unless otherwise stated:
$\log{N_\mathrm{H}^{\mathrm{Torus}}}/{\mathrm{cm}}^{-2} = 24.0$, $\sigma = 20.0^\circ$, $\log{N_\mathrm{H}^\mathrm{Polar}}/{\mathrm{cm}}^{-2} = 21.5$,  $\Delta_{\mathrm{in}} = 10.0^\circ $, $Y_{\mathrm{Polar}} = 200$,
$\Gamma = 1.8$, $E_{\mathrm{cut}} = 370 ~\mathrm{keV}$.
Figure~\ref{figure:scattered} displays the images of the reflection
components around 0.5 $\mathrm{keV}$, 1.7 $\mathrm{keV}$ (including Si
$\mathrm{K}\alpha$), 3.0 $\mathrm{keV}$, 6.4 $\mathrm{keV}$ (including Fe
$\mathrm{K}\alpha$), and 20.0 $\mathrm{keV}$ when the angle of
inclination changes to $0^{\circ}$, $30^{\circ}$, $60^{\circ}$,
$90^{\circ}$.
As noticed, the emission around Si $\mathrm{K}\alpha$
(1.7 keV) is dominated by that from of the polar dusty gas
at high inclination angles. By
contrast, the emission around Fe $\mathrm{K}\alpha$ (6.4 keV) comes
almost entirely from the torus. These results support the suggestion
by \citet{2019MNRAS.490.4344L} that Si $\mathrm{K}\alpha$ / Fe
$\mathrm{K}\alpha$ is a good indicator to search for objects with
significant polar dusty gas.
  Figure~\ref{figure:model_i10} plots the total spectra including the
  direct component and the reflection components from the torus and
  polar dusty gas for $i=10^\circ$ (left) and $i=80^\circ$ (right).  As
  noticed, in the face-on case, the direct component 
  dominates over the reflected component, so that any enhancement of
  the polar dust reflection would be hardly detectable in actual
  observations. Hence, we hereafter focus on the edge-on case.

\begin{figure*}
\centering
\includegraphics[width=0.45\textwidth]{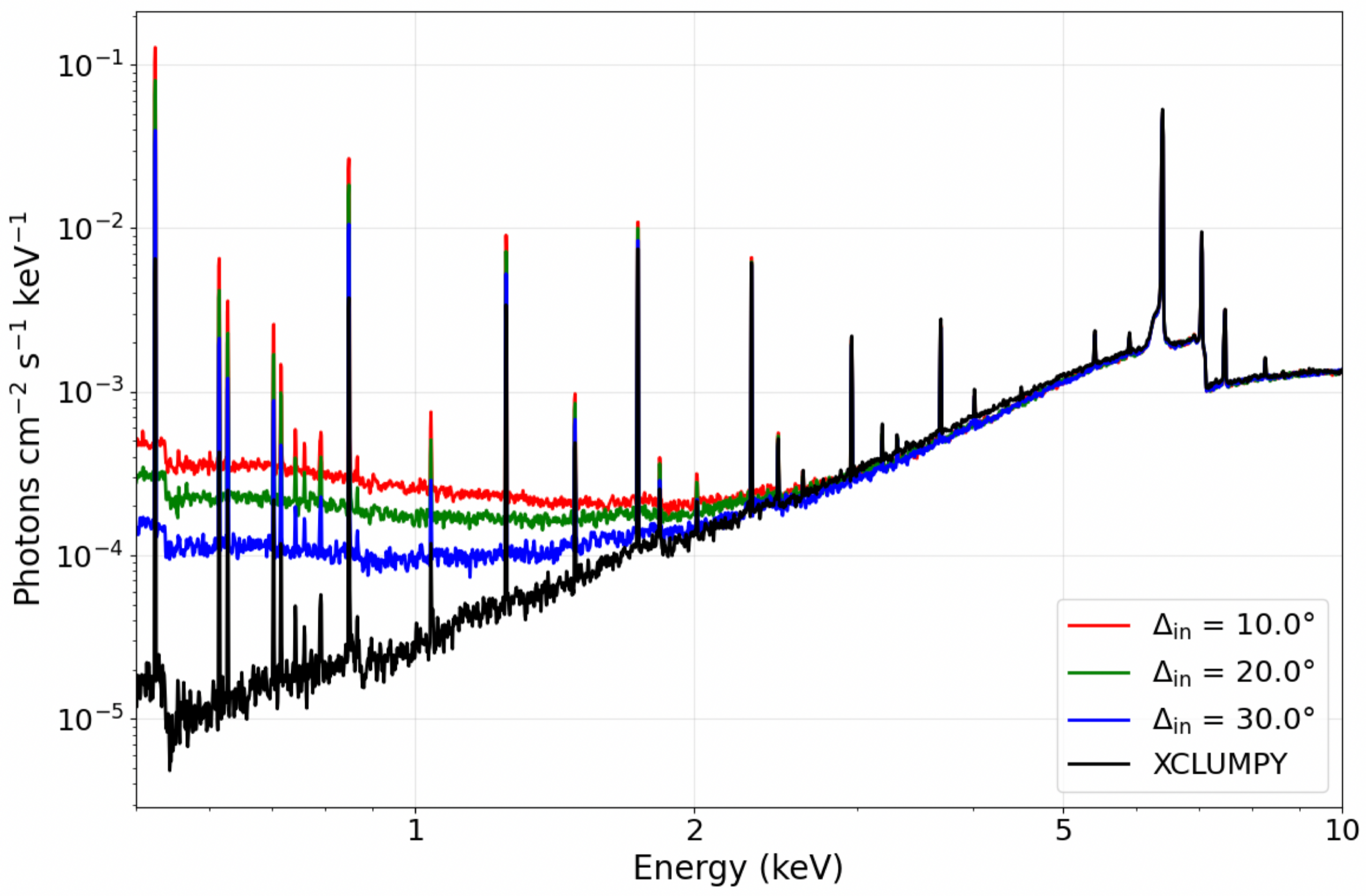}
\includegraphics[width=0.45\textwidth]{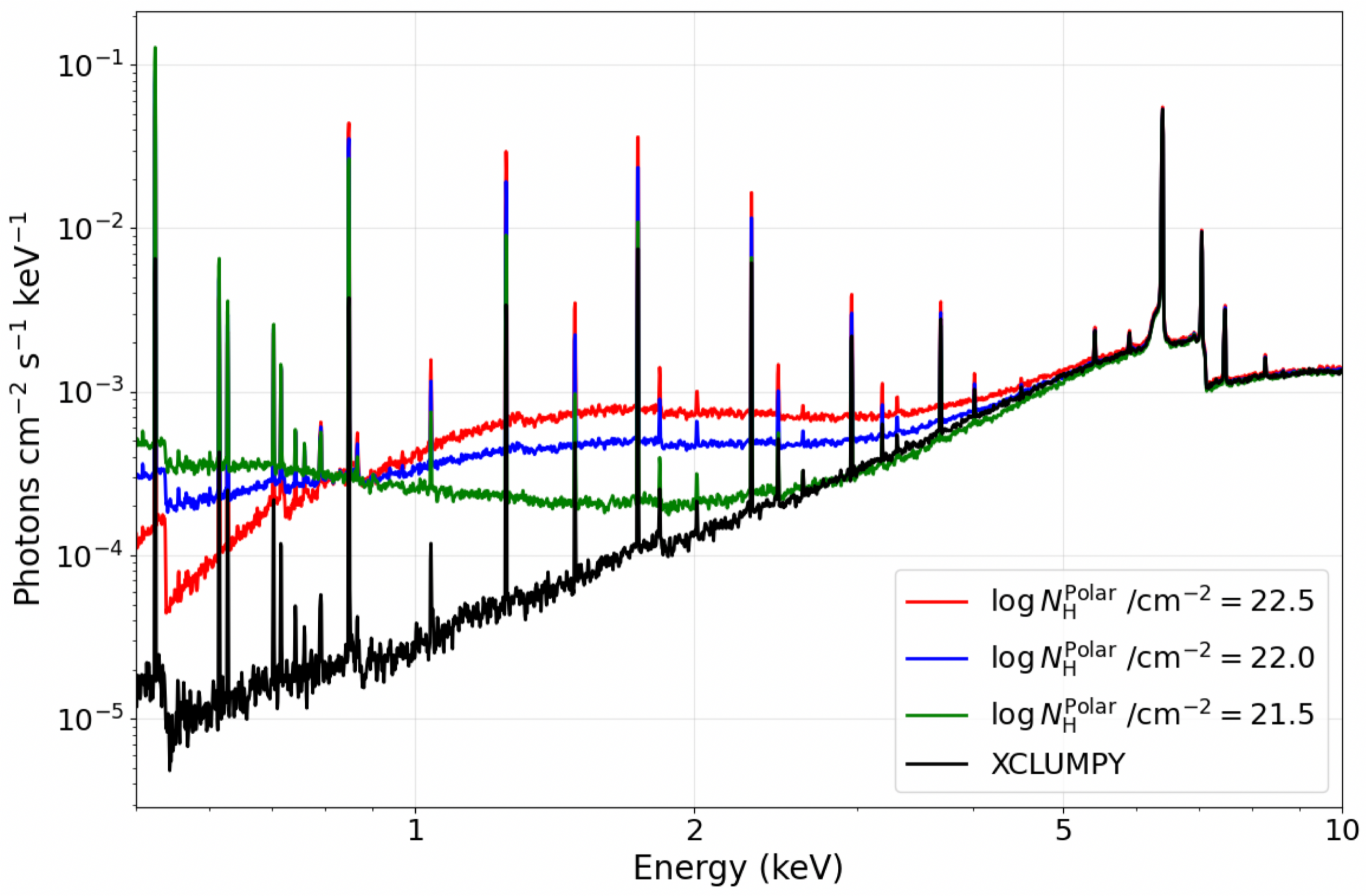}
\caption{
  (Left) The X-ray reflection spectra observed edge-on ($i=80^\circ$) calculated for different half
  opening angles of the polar dusty gas.
  The
  red, green, and blue curves correspond to ${\Delta}_\mathrm{{in}} =
  10.0^\circ,~20.0^\circ$, and $30.0^\circ$, respectively, and the
  black one to the torus-only case. We assume $\log{N_\mathrm{H}^\mathrm{Polar}}/{\mathrm{cm}}^{-2} = 21.5$, $Y_{\mathrm{Polar}} = 200$, $\Gamma =
  1.8$, and $E_{\mathrm{cut}} = 370 ~\mathrm{keV}$.
  (Right) Those for different effective hydrogen column densities of the polar dusty gas.
  The red, green, and blue curves correspond to $\log{N_\mathrm{H}^\mathrm{Polar}/{\mathrm{cm}}^{-2}} = 22.5, 22.0,$ and $21.5$  respectively, and the
  black one to the torus-only case. We assume
  ${\Delta}_\mathrm{{in}} = 10.0^\circ$. The other parameters are the same as in the left panel.
 \label{figure:nhpo}
 \label{figure:hoa}}
\end{figure*}

Figure~\ref{figure:hoa} (left)
plots the reflection spectra observed edge-on $(i = 80^\circ)$ with different 
half opening angles $\Delta_{\mathrm{in}}$. For comparison,
the reflection spectrum only from the torus (without polar dusty gas),
calculated using the XCLUMPY(SKIRT) model (Appendix~\ref{section:XCLUMPY-SKIRT}), is also
plotted.
As noticed, the flux at low energies ($<$5
keV) becomes larger when the polar dusty gas is present, confirming
the results by \citet{2019MNRAS.490.4344L} and
\citet{2022MNRAS.512.2961M}. The soft X-ray flux ($<$5 keV) decreases with
$\Delta_{\mathrm{in}}$.
This can be understood as follows.
For the default parameter set, the polar dusty gas is optically thin.
For photoelectric absorption at $>$0.5 keV, it basically produces
unabsorbed scattered component by keeping the shape of the intrinsic
power law component. The amount of this additional component is
roughly proportional to the total number of electrons (hence hydrogen atoms)
in the polar dusty gas that is not blocked by the optically thick torus when viewed from the central engine; hence, the soft X-ray flux decreases with $\Delta_{\mathrm{in}}$ as
the torus obscuration becomes more significant.
Furthermore, a smaller $\Delta_{\mathrm{in}}$ of the polar dust increases the solid angle subtended by the polar dust as viewed from the line of sight, which also contributes to the stronger reflected spectrum.
Figure~\ref{figure:nhpo} (right) shows the dependence of the reflected X-ray
spectrum at $i = 80^\circ$ on the hydrogen column density of the polar dusty gas ($N_\mathrm{H}^\mathrm{Polar})$.
The contribution of scattered X-rays from
the polar dusty gas increases with $N_\mathrm{H}^\mathrm{Polar}$ in this range. It is notable that, as the density increases, the soft X-ray flux becomes more attenuated, due to self-absorption in the polar dusty gas. The trend is similar to that reported by 
\citet{2019MNRAS.490.4344L}.

\begin{figure*}
\centering
\includegraphics[width=0.46\textwidth]{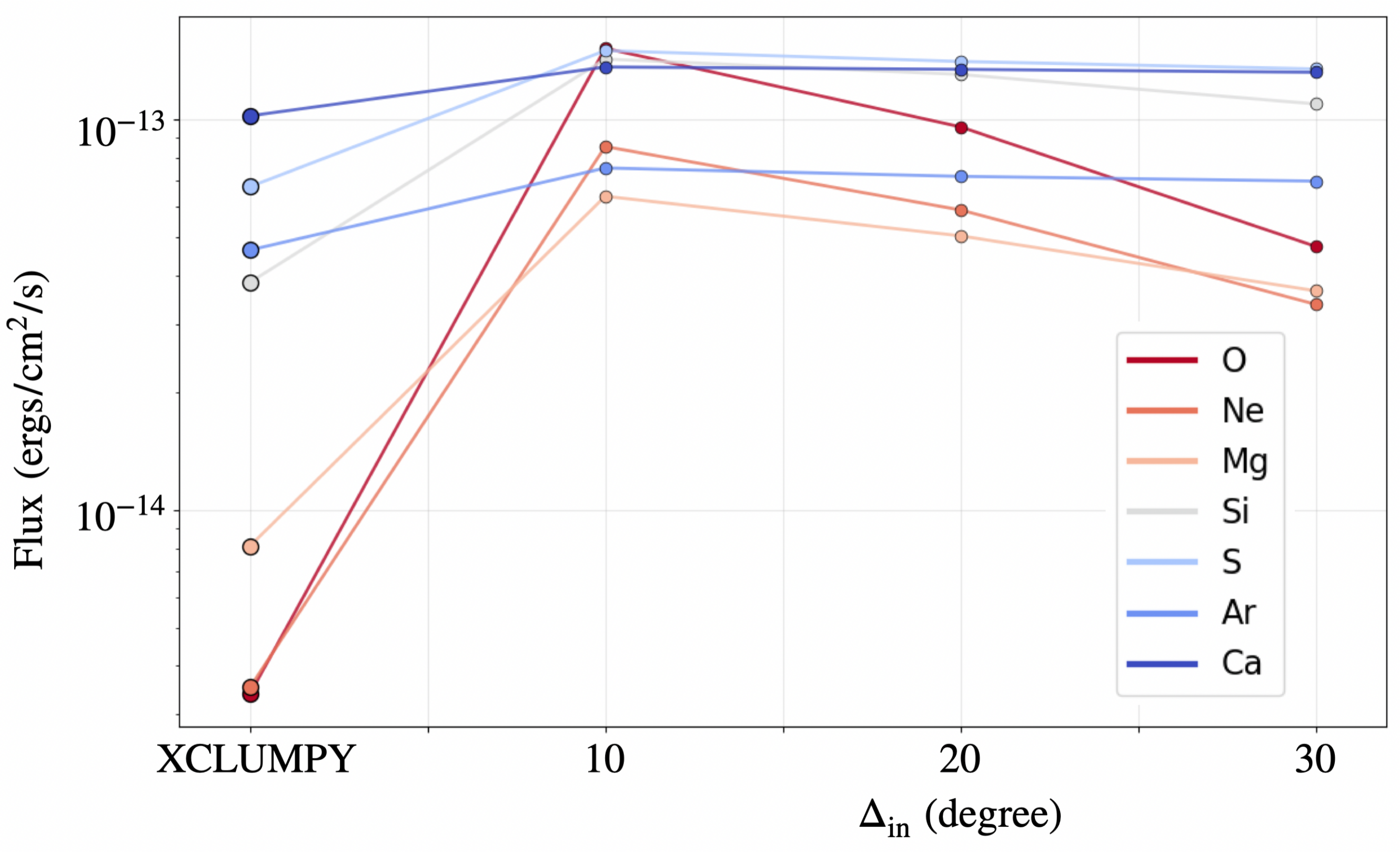}
\includegraphics[width=0.46\textwidth]{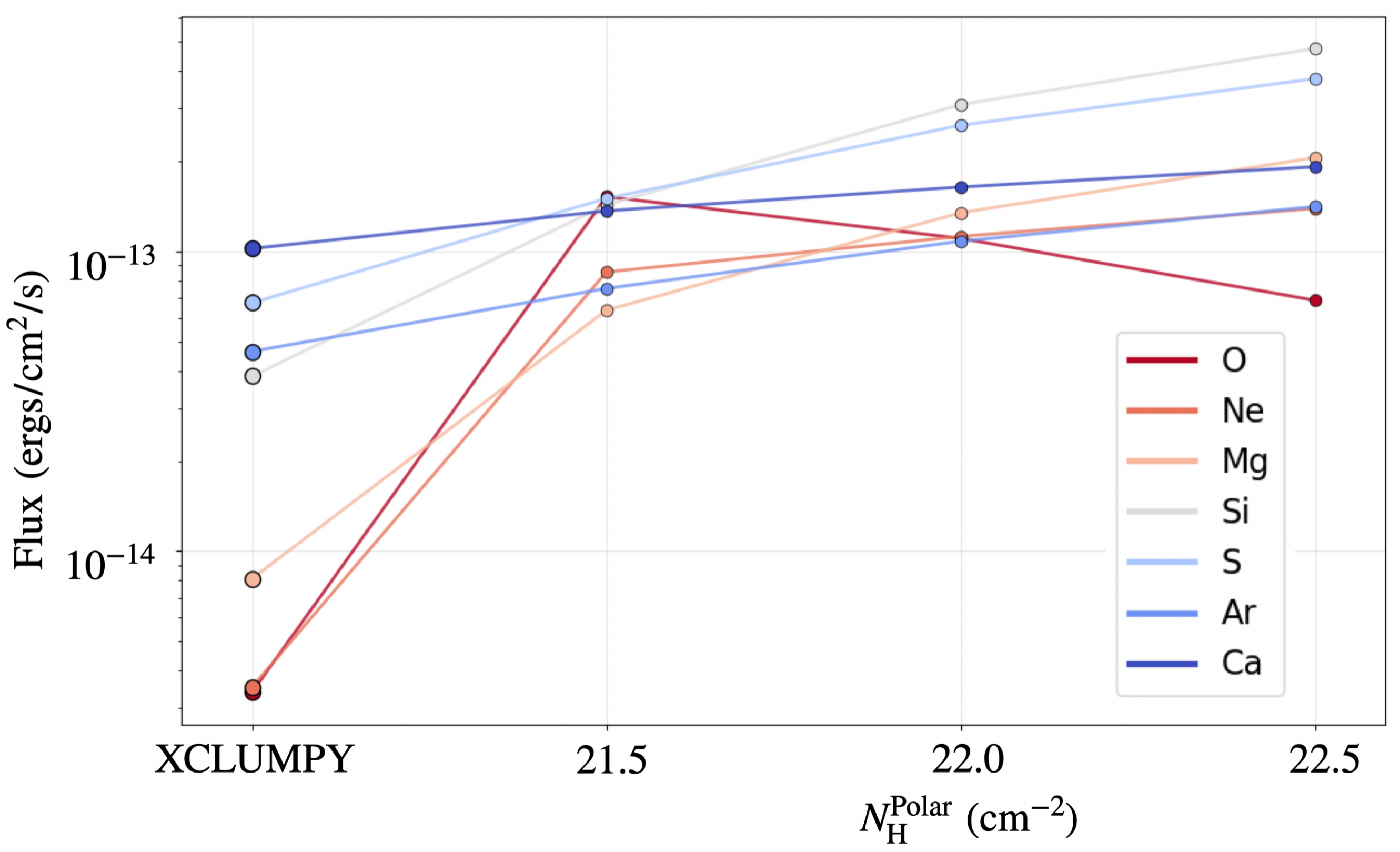}
\caption{
(Left) The fluxes of fluorescent lines of O, Ne, Mg, Si, S, Ar, and Ca for different $\Delta_{\mathrm{in}}$. The leftmost one corresponds to the torus-only case.
  (Right) Those for different hydrogen column densities of the polar dusty gas.
  The leftmost one corresponds to the torus-only case. In both panels, we assume the default parameter set and $i=80^\circ$.
 \label{figure:line_hoa}
 \label{figure:line_nhpo}}
\end{figure*}

We investigate the relation between the fluxes 
of fluorescent emission lines 
and the polar-dusty-gas parameters in Figure~\ref{figure:line_hoa}.
Here we consider fluorescent lines of O, Ne, Mg, Si, S, Ar, and Ca, because the effect by the polar dusty gas is important at low energies ($<$4 keV) as shown in Figure~\ref{figure:nhpo}.
As noticed, the line fluxes in the torus plus polar dusty gas case
are always larger than those in the torus-only case. They increase with decreasing $\Delta_{\mathrm{in}}$, and with increasing $N_\mathrm{H}^\mathrm{Polar}$ except for oxygen and neon in the adopted range of $N_\mathrm{H}^\mathrm{Polar}$ values. This is because, similarly to the Compton scattered continuum, the fluorescent-line flux from the polar dusty gas should be roughly proportional to the total number of atoms in the volume irradiated by the intrinsic X-ray emission (i.e., that not blocked by the torus) unless the gas is optically thick for photoelectric absorption. In the high  column  density case ($N_\mathrm{H}^\mathrm{Polar}/{\mathrm{cm}}^{-2} = 22.5$), self-absorption within the polar dusty gas reduces the observed flux of the oxygen line (0.5 keV) and neon line (0.8 keV).

\section{Application to NGC 4388}\label{section:NGC4388}

As an example, we apply the IMPACT and IMPACTX models to NGC 4388, one
of the closest ($z=0.00842$) Seyfert 2 galaxies \citep{1982AJ.....87.1628H}
located at a distance of 18.0 Mpc \citep{2014MNRAS.444..527S}. This
object is a member of the sample consisting of 28 AGNs analyzed by
\citet{2021ApJ...906...84O} and Paper~I, for which nuclear infrared SEDs obtained by
high angular-resolution mid-infrared imaging are available
\citep{2019MNRAS.486.4917G}. It is X-ray bright, showing the brightest Fe K$\alpha$
line among Compton-thin obscured AGNs \citep{2011ApJ...727...19F}, and is a target
of a XRISM observation in GO-1 (PI: Ueda). Thus, NGC 4388 is an ideal
target to demonstrate the utility of our model.

\ifnum0=1

\begin{deluxetable*}{lccccccccccccc}
\tablewidth{\textwidth}
\tablecaption{Properties of NGC 4388\label{tab-obj}}
\tablehead{
Object & Right ascension & Declination & Redshift & Class & Group & $\log M_\mathrm{BH}/M_\mathrm{\odot}$ & $\log L_\mathrm{X}$ & $\log \lambda_\mathrm{Edd}$ & X-ray Ref. & $M_\mathrm{BH}$ Ref.  \\
(1) & (2) & (3) & (4) & (5) & (6) & (7) & (8) & (9) & (10) & (11)
}
\startdata
NGC 4388 & 12$^\mathrm{h}$25$^\mathrm{m}$46$^\mathrm{s}$.747 & $+$12$^\mathrm{d}$39$^\mathrm{m}$43$^\mathrm{s}$.51 & 0.00842 & 2 & Obscured & 8.00 & 43.0 & -1.84 & a & b \\
\enddata
\tablecomments{
(1) Galaxy name. (2)--(3) Position (J2000) from the NASA/IPAC extragalactic database (NED). (4) Redshift from the NED. (5) Optical AGN classification from the NED. 
(6) Sub-group of AGNs based on X-ray obscuration. (7) Logarithmic black hole mass. (8) Logarithmic X-ray luminosity (2--10 keV). (9) Logarithmic Eddington ratio
(10) Reference for the X-ray results delivered by the XCLUMPY model. (11) Reference for the black hole mass.}
\tablerefs{(a):\citealt{2021ApJ...906...84O}, (b):\citealt{2016ApJS..225...14K}}
\end{deluxetable*} 
\fi


We analyze the infrared SED of the nucleus compiled in Paper~I,
covering the 1--100 $\mu$m band.
It consists of the photometric or spectroscopic data summarized in Table~\ref{table:SED}. Figure~\ref{figure:IR_fit} plots the IR SED (including upper limits) as a function of wavelength. 
We utilize the same broadband X-ray spectrum reduced by \citet{2021ApJ...906...84O}. It
consists of the Suzaku/XIS and NuSTAR spectra observed in 2005 and
2013, covering the 0.5-10 keV and 8--75 keV bands, respectively.  The
observed spectra folded with the energy response are plotted in
Figure~\ref{figure:broad} (a). The observation log is given in Table~\ref{tab-obj}.

\subsection{Spectral Analysis}

In order to determine the AGN structure including the torus and polar
dusty gas, the simplest way is to perform simultaneous fitting to the
IR and X-ray spectra by leaving all geometrical parameters free at
once. In practice, however, since the number of data points of the IR
SED are much smaller than that of the X-ray spectra in our case, the
obtained results may heavily depend on the X-ray data and constraints
from the IR data could not be well reflected. To avoid this situation,
we take a step-by-step approach as described below. We recall that a
large fraction of IR emission from an obscured AGN comes from the
polar dusty gas (e.g. \citealt{2006A&A...452..459H,2012ApJ...755..149H,2013ApJ...771...87H};
\citealt{2014A&A...563A..82T};
\citealt{2014A&A...565A..71L,2016A&A...591A..47L};
\citealt{2018ApJ...862...17L}; and \citealt{2019MNRAS.489.2177A}), whereas X-ray reflection component is dominated
by that from the torus region except for extreme values of the number
density and half opening angle (Figures~\ref{figure:scattered} and  \ref{figure:nhpo}).

\begin{enumerate}

\item First we fit the X-ray spectra (0.5--75 keV) by ignoring the
  polar dusty gas, utilizing the XCLUMPY(SKIRT) model (Appendix~\ref{section:XCLUMPY-SKIRT}).

\item Next we fit the IR SED (1--100 $\mu$m) with the IMPACT model to
  constrain the polar-dusty-gas parameters by limiting the range of
  torus parameters within uncertainties estimated in the first step.

\item Finally, we fit the IR and X-ray spectra simultaneously with the
  IMPACT and IMPACTX models, respectively. Here we fix the parameters related to the polar dusty gas ($\Delta_{\mathrm{in}}$, $Y_{\mathrm{Polar}}$, and $N_\mathrm{H}^\mathrm{Polar}$) at the best-fit values.

\end{enumerate}

For the fitting, we utilize the XSPEC package (version 12.15.0) both for
the IR and X-ray spectral analysis, by converting the units of \AA$~$to keV in the IR spectrum. In our analysis, we fix
$E_{\mathrm{cut}} = 370 ~\mathrm{keV}$ and $i = 70^{\circ}$ from \citealt{2021ApJ...906...84O}, which are not well constrained from our data.
We also limit $\Delta_{\rm in} > 20^{\circ}$, to make it consistent with the observed opening angle of the ionization cone in NGC 4388 \citep{2017MNRAS.469.2720G}.

\subsubsection{IR SED Model} \label{IR_SED_fitting}

The model use to fit the IR SED is represented as follows in the XSPEC
terminology:
\begin{eqnarray}
\mathrm{model_{IR}}
&=& \textsf{zdust}*\textsf{zdust} \nonumber\\
&*&  \textsf{(atable\{impact\_disk\_direct.fits\}} \nonumber\\
&+&  \textsf{atable\{impact\_disk\_scattered.fits\}} \nonumber\\
&+&  \textsf{atable\{impact\_dust\_direct.fits\}} \nonumber\\
&+&  \textsf{atable\{impact\_dust\_scattered.fits\})} 
\end{eqnarray}

The two \textsf{zdust} terms represent dust extinction in our Galaxy and the host galaxy. The color excess $E(B - V)$ in the former is fixed at the total Galactic value determined from the dust map \citep{1998ApJ...500..525S}, whereas that in  the latter is left free by adopting the extinction curve of Small Magellanic Cloud (SMC; \citealt{1984A&A...132..389P}), which is
preferred to represent that in an AGN
(e.g. \citealt{2004AJ....128.1112H, 2009ApJ...690.1250S, 2012MNRAS.427.3103B, 2021A&A...654A..93B, 2023ApJS..265...37Y}). The first two \textsf{atable} terms
represent the direct and scattered emission from the accretion disk,
and the third and fourth ones the direct and scattered emission 
from the torus plus polar dusty gas. The geometrical parameters
in these tables are all linked together.

\subsubsection{X-ray Spectral Model}\label{section:X-fit}


The X-ray spectral model is represented as follows in the XSPEC terminology:
\begin{eqnarray}\label{modelX}
\mathrm{model_X}
&=& \textsf{const1*phabs} \nonumber\\
&*&  \textsf{(const2*zphabs*cabs*zcutoffpl } \nonumber\\
&+& \textsf{const3*zcutoffpl + atable\{impactx.fits\} }\nonumber\\
&+& \textsf{zgauss + apec1 + apec2)}\nonumber
\end{eqnarray}

\begin{enumerate}
\item The \textsf{const1} term is a cross-normalization constant to adjust small differences in the absolute flux calibration among different instruments. We set those of Suzaku/FIXIS and NuSTAR/FPM to unity as references. The \textsf{phabs} term represents the Galactic absorption, whose hydrogen column density is fixed at $2.87 \times {10}^{20}$ $\mathrm{cm}^{-2}$, a value estimated by the method of \citet{2013MNRAS.431..394W}.
  
\item The first term represents the transmitted component absorbed by the torus. The line-of-sight column density is determined by the torus parameters with this equation
\begin{equation} 
 N^{\mathrm{LOS}}_\mathrm{H} = N^{\mathrm{Equ}}_\mathrm{H} \exp{\biggl (- \frac{{(\theta - \pi/2)}^2}{\sigma^2} \biggr )}.
\end{equation}
Note that the cabs model assumes free-electron scattering, whereas this work (SKIRT) assumes scattering by electrons bound to atoms. We confirm the difference in the total scattering cross section ($<10\%$) does not affect our fitting results.

\item The second one is an unabsorbed scattered component from ionized gas in the polar region (i.e., that is not account for in the IMPACTX model). The \textsf{const2} term ($C_{\mathrm{TIME}}$) is a constant to consider time variability between the Suzaku and NuSTAR observations. We do not multiply this constant to the scattered component and the reflection component. This is because the sizes of the scatterer and reflector are likely larger than parsec scales and hence little time variability is expected among the two observations (2005 and 2013). The normalization of the second term is linked to those in the first team so that the \textsf{const3} term denotes the scattered fraction.
  
\item The third term represents the continuum and fluorescent lines in the reflection component from the torus plus polar dusty gas (IMPACTX). The normalization and photon index are linked to those in the first term (\textsf{zcutoffpl}).

\item The \textsf{zgauss} term represents Fe K$\alpha$ fluorescent line from the broad line region (BLR).
 Recent high-resolution observations with XRISM suggest that the Fe K$\alpha$ emission line consists of multiple components, including a narrow line  originating from the torus and a broader line potentially originating from the BLR (\citealt{2024ApJ...973L..25X, 2025arXiv250702195B}). The equivalent width and Doppler velocity width of this component are fixed at 50 eV and $5.4\times10^3$ km s$^{-1}$ (FWHM), following the XRISM result on NGC 4388 (Fujiwara et al., in preparation).  

\item The two \textsf{apec} terms represent optically-thin thermal emission from the host galaxy.

\end{enumerate}

\subsection{Fitting Results and Discussion}

\begin{figure}[htb]
\centering
\begin{minipage}[h]{1.0\columnwidth}
    \centering
    \includegraphics[width=1 \columnwidth]{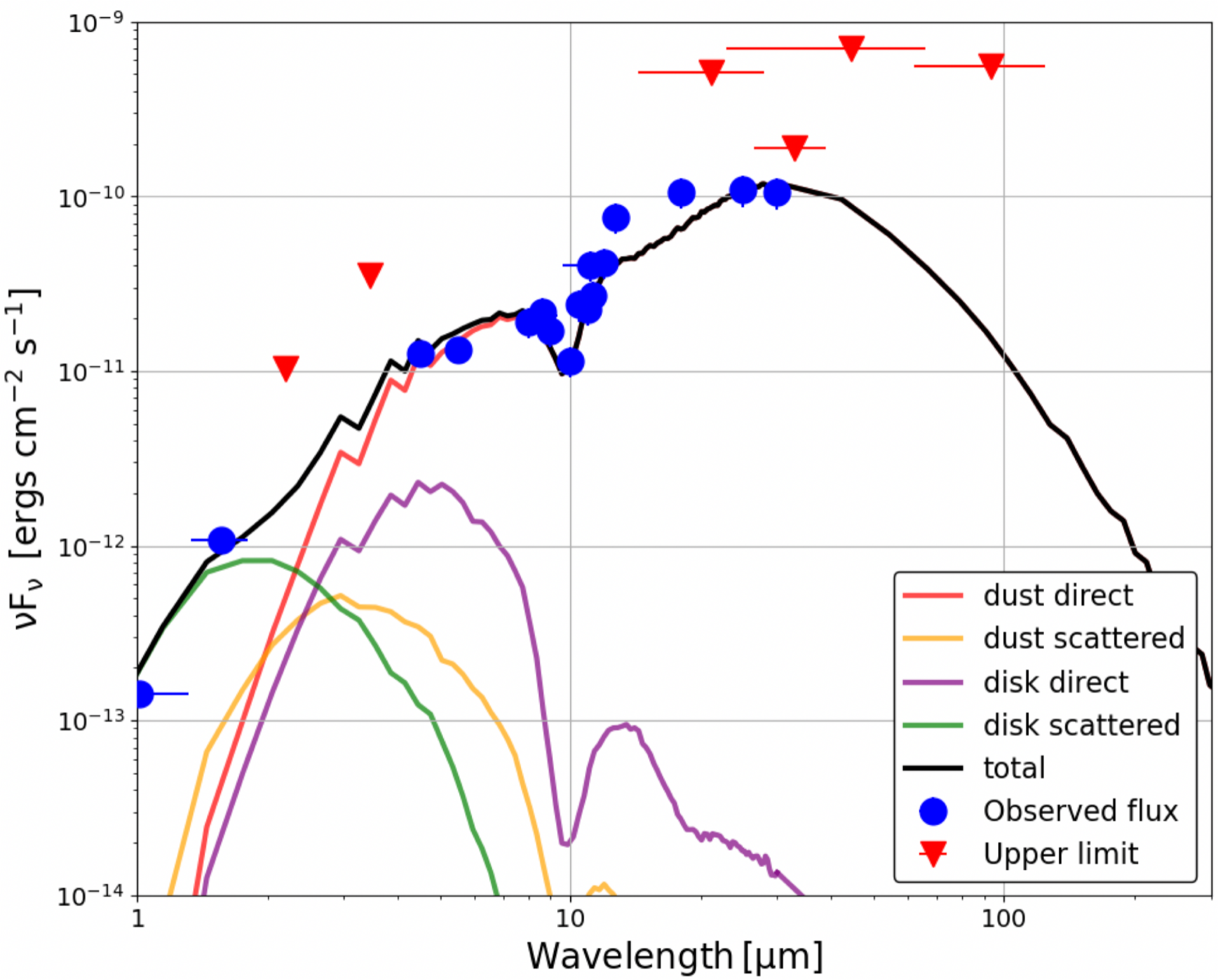}
\end{minipage}
\caption{The infrared SED and the best-fit IMPACT model. The black line is the total model. The red and yellow curves are the direct emission from the torus plus polar dusty gas and its scattered emission, respectively. 
  The purple and green ones are the direct emission from the accretion disk and its scattered emission, respectively.
}
\label{figure:IR_fit}
\end{figure}

\begin{figure*}[htb]
\centering
\includegraphics[width=0.48\textwidth]{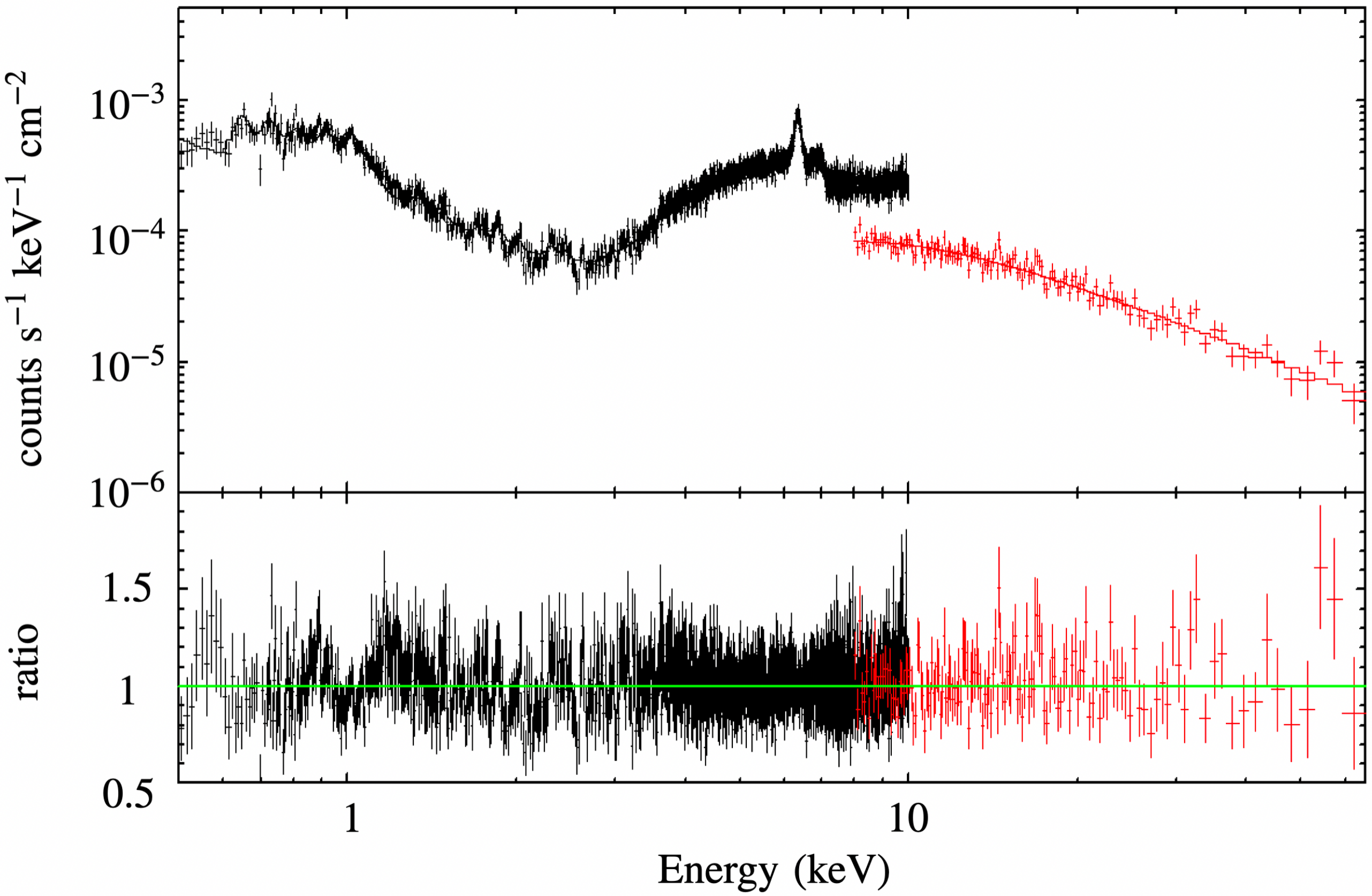}
\includegraphics[width=0.48\textwidth]{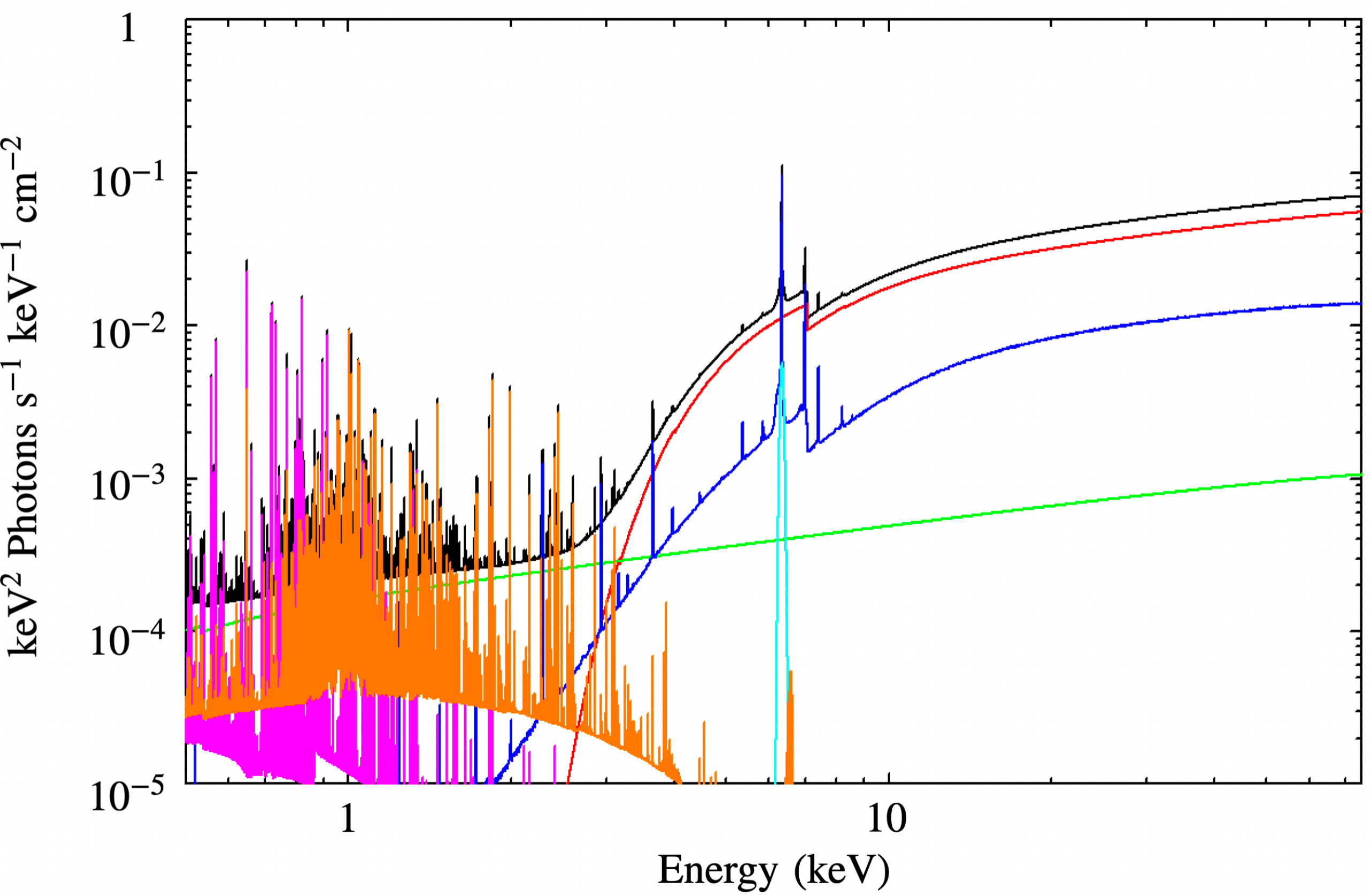} 
\caption{(a) The broadband X-ray spectra of NGC 4388 folded with the energy response  (black:
  Suzaku/XIS, red: NuSTAR/FPMs). The best-fit model is overplotted. 
  (b) The best-fit model in units of $E F_E$
  (black: total,
  red: transmitted component,
  green: scattered component from ionized gas,
  blue: reflection component in the IMPACTX model,
  light blue: emission line from the BLR,
  orange and purple: emission from optically-thin thermal plasma.)
  }
\label{figure:broad}
\end{figure*}

\begin{deluxetable*}{lcCCcc}[htb]
  \tablecaption{Best fit parameters of the IR SED and X-ray spectra of NGC 4388}
 \label{table:broadbest}
 \tablewidth{0pt}
 \tablehead{
 \colhead{Region} & \colhead{No.} & \colhead{Parameter} & \colhead{Best Fit Value} & \colhead{Units}
 }
 \startdata
 Torus&(1) & $\log{N_\mathrm{H}^\mathrm{Equ}}$ & $24.5^{+0.0}_{-0.1}$ & $\mathrm{cm}^{-2}$   \\
        &(2) & $\sigma$ & $13.5^{+1.2}_{-0.1}$ & degree \\
        &(3) & $i$ & $70.0^a$ & degree \\
        &(4) & $Z(\mathrm{Fe})$ & $2.05^{+0.14}_{-0.14}$ & solar \\
        &(5) & $\Gamma_{\mathrm{Suzaku}}$ & $1.49^{+0.03}_{-0.04}$ &  \\
        &(6) & $\Gamma_{\mathrm{NuSTAR}}$ & $1.72^{+0.03}_{-0.04}$ &  \\
        &(7) & $E_{\mathrm{cut}}$ & $370^a$ & keV  \\
        &(8) & $N_X$ & $1.10^{+0.05}_{-0.09}\times{10^{-2}}$ &$\mathrm{photons}~\mathrm{keV}^{-1}~\mathrm{cm}^{-2}~\mathrm{s}^{-1}$\\
        &(9) & $f_{\mathrm{scat}}$ & $1.27^{+0.10}_{-0.04}\times{10}^{-2}$ &   \\
       Polar dusty gas$^b$&(10) & $\Delta_\mathrm{in}$ & $20.0^{+0.7}_{-0.0}$ & degree\\
       &(11) & $Y_{\mathrm{Polar}}$& $6.31^{+0.64}_{-0.97}\times{10}^{2}$ &  \\
       &(12) &$\log{N_\mathrm{H}^\mathrm{Polar}}$ & $23.0^{+0.0}_{-0.1}$& $\mathrm{cm}^{-2}$ \\
       Others&(13) & $T_{\mathrm{apec1}}$ & $1.20^{+0.05}_{-0.03}$ & keV \\
        &(14) & $T_{\mathrm{apec2}}$ & $0.31^{+0.02}_{-0.02}$ & keV &\\
        &(15) & $N_{\mathrm{apec1}}$& $1.84^{+0.18}_{-0.12}\times{10^{-4}}$ &  \\
        &(16) & $N_{\mathrm{apec2}}$ & $1.46^{+0.15}_{-0.15}\times{10^{-4}}$ &  \\
        &(17) & $C_{\mathrm{TIME}}$ & $0.58^{+0.08}_{-0.07}$ & \\
        &(18) & $E(B-V)$& $1.96^{+0.25}_{-0.23}$& mag\\
          &  & $\chi^2/\mathrm{dof}$ & $2080.70/1996$ &  \\
 \enddata
 \tablecomments{
 (1) Hydrogen column density along the equatorial plane. 
 (2) Torus angular width. (3) Inclination angle. (4) Abundance of iron relative to hydrogen. (5)
 Photon index for Suzaku data. (6) Photon index for NuSTAR data. (7) Cutoff energy. (8) Normalization of the intrinsic power law component at 1 keV. (9) Scattering fraction.
   (10) Half opening angle of the polar dusty gas.
   (11) Radial thickness parameter of the polar dusty gas ($R_{\mathrm{Polar}}/r_{\mathrm{in}}$).
   (12) Hydrogen column density of the polar dusty gas. (13), (14) Temperature of optically thin clouds. (15), (16) Normalization of the optically thin components. (17) Time variability constant. (18) Dust extinction in the host galaxy.
 } 
 \tablenotetext{a}{The parameter is fixed.}
 \tablenotetext{b}{The errors are obtained from the fitting of the IR SED (the second step in Section~4.1).}
\end{deluxetable*}

Our model can reproduce both the IR and broadband X-ray spectra of NGC
4388 reasonably well (within a factor of 2 for the IR SED). Table~ \ref{table:broadbest}
summarizes the best-fit parameters; the errors on the parameters of
the polar dusty gas are those derived in the second step in
Section~\ref{IR_SED_fitting}. The best-fit models in the IR and X-ray bands are
overplotted in Figures~\ref{figure:IR_fit} and \ref{figure:broad},
respectively. Figure~\ref{figure:broad} (b) displays the best-fit X-ray spectral model
in units of $E F_E$ ($F_E$ is the energy flux density) with
contributions from different components.

We find that the polar dusty gas in NGC 4388 has a large hydrogen column density along the radial direction ($\log{N_\mathrm{H}^\mathrm{Polar}/\mathrm{cm}^{-2}} = 23.0^{+0.0}_{-0.1}$), which are
constrained by the IR SED.
The torus parameters, $\sigma = 13.5^{+1.2}_{-0.1}$ degree and 
  $\log{N_\mathrm{H}^\mathrm{Equ}/{\rm cm^{-2}}} =  24.5^{+0.0}_{-0.1}$.

We find that the line-of-sight absorption toward the
central engine is Compton thin ($N_\mathrm{H}^{\mathrm{LOS}}/{\rm cm^{-2}} =  23.5^{+0.1}_{-0.1}$), confirming the
previous results (\citealt{1990MNRAS.242..262H};
\citealt{1997MNRAS.285..683I}; \citealt{1999ApJ...523..521F};
\citealt{2002ApJ...571..234R}; \citealt{2003MNRAS.345..369I};
\citealt{2004ApJ...614..641B}; \citealt{2008PASJ...60S.263S};
\citealt{2017ApJ...843...89K}; \citealt{2011ApJ...727...19F}; \citealt{2019ApJ...884..106M};
\citealt{2021ApJ...906...84O}).

\begin{figure}[htb]
\centering
\begin{minipage}[h]{1.0\columnwidth}
    \centering
    \includegraphics[width=1.0 \columnwidth]{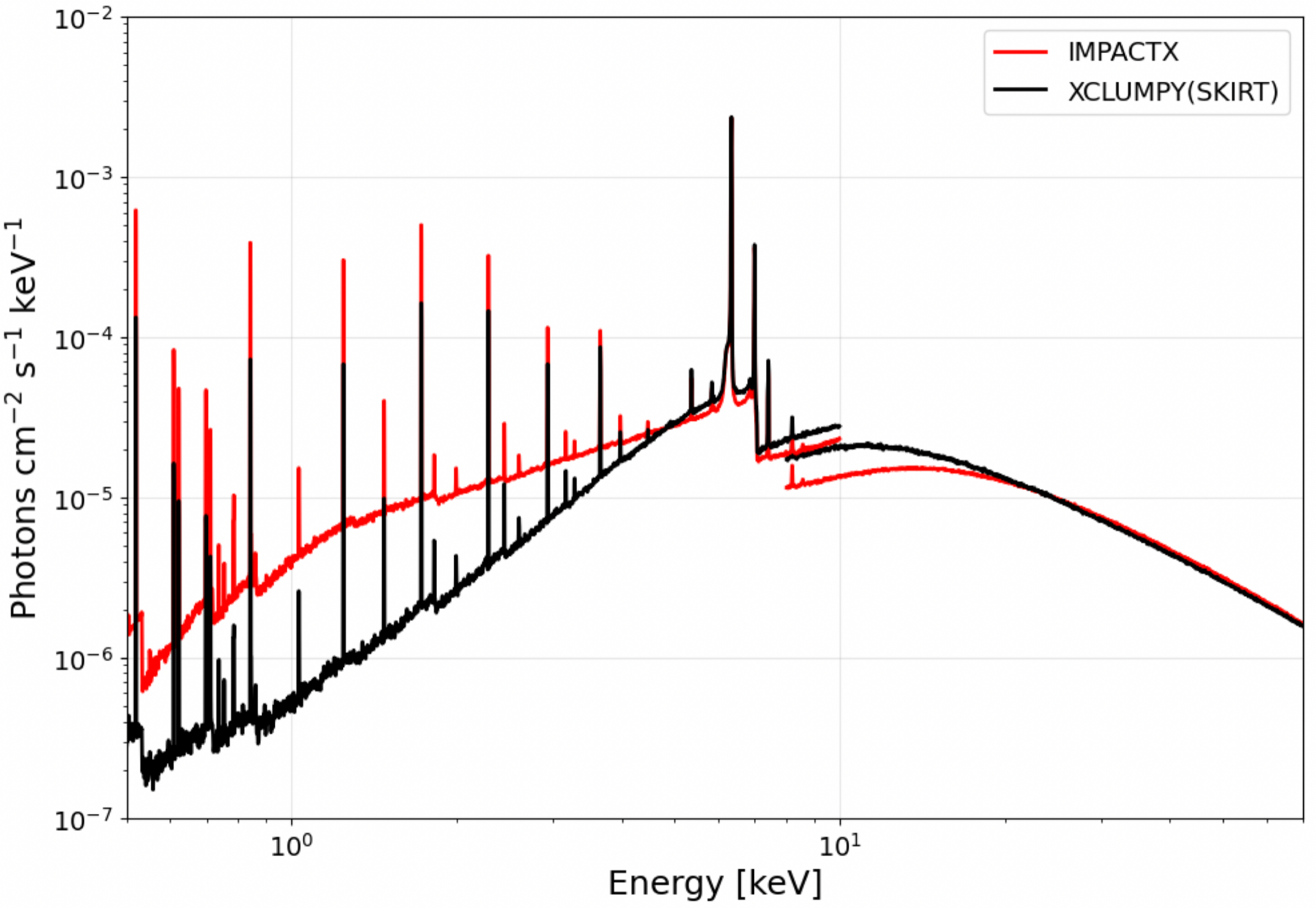}
\end{minipage}
\caption{
  Comparison of best fit X-ray spectral models of NGC 4388 obtained with IMPACTX/IMPACT models (red)
  and with XCLUMPY(SKIRT) model (black). 
}
\label{figure:ngc4388_diff}
\end{figure}

Figure \ref{figure:ngc4388_diff} compares the best-fit model obtained
with IMPACTX (with polar dusty gas) and that with XCLUMPY(SKIRT)
(without it). As noted, they are significantly different; the former
shows an excess around 2 keV due to the reflected spectrum from the
polar dusty gas. The reason why both models can reproduce the whole
X-ray spectra despite of this difference is that the excess flux from
the polar dusty gas can be accounted for by increasing the fraction of
the ``scattered component'' from ionized gas (the 2nd term in equation
\ref{modelX}). This implies that it is difficult to accurately determine the
reflection component from the polar-dusty-gas using X-ray alone and
complementary IR SED data are necessary.
Assuming our best-fit geometry, we estimate the covering fraction by
the torus and/or polar dusty gas with $\log N_{\mathrm{H}}/\mathrm{cm}^{-2} > 22$ to be 0.67,
which is similar to that estimated from the torus-only case, 0.67,
calculated according to equation 6 in \citet{2021ApJ...906...84O}.
This value is also consistent with the relation by \citet{2017Natur.549..488R} at the
Eddington ratio of $\log \lambda_{\rm Edd} = -1.84$ \citep{2021ApJ...906...84O}, implying that
the polar dusty gas is also responsible for obscuration of an AGN. 

To summarize, we demonstrate that simultaneous analysis of the IR and
X-ray spectra is very useful to constrain the nuclear structure
including the torus and polar dusty gas. The contribution from the
polar dusty gas in the X-ray spectrum can be significant, in
particular when the density is high and/or opening angle is small.
Hence, as pointed out by \citet{2019MNRAS.490.4344L}, we have a chance to measure the
outflow velocity of the polar dusty gas by detecting blue-shifted
fluorescent lines in high energy resolution spectra. The torus geometry
determined by assuming the torus-only geometry, as done in the
previous works, are not largely affected by including the
polar-dusty-gas component. However, it is possible that the covering
fraction by surrounding gas may have been underestimated in objects
with strong dusty outflow.

\section{Conclusion}\label{conclusion}

\begin{enumerate}

\item We have constructed the IMPACTX model, a generic X-ray spectral
  model for the reflection components from the clumpy torus and polar
  dusty gas in an AGN, where the same geometry as in the IMPACT model
  is adopted. This enables us to directly compare the X-ray and
  infrared results, which trace the distribution of all matter and
  heated dust, respectively.

\item  We have confirmed that the polar dusty gas contributes to the low energy side $(< 5~\mathrm{keV})$ of the X-ray spectrum, including the scattered continuum and fluorescent emission lines. Their fluxes increase with the density and decreases with the half-opening angle unless self absorption within the polar dusty gas becomes important.

\item As an example, we simultaneously analyze the IR SED and
  broadband X-ray spectra of NGC 4388 with the IMPACT and IMPACTX
  models, respectively. We have found a solution that reproduces both
  data reasonably well. Comparing with the result obtained by the torus-only 
  geometry, our new model predicts a significantly higher flux around 2 keV due to the polar dusty gas, which is difficult to separate by using X-ray data alone.  This demonstrates the importance of using both IR and X-ray data to constrain the nuclear structure. 

\end{enumerate}

This work was supported by the Japan Society for the Promotion of Science (JSPS) KAKENHI grant number 20H01946 (Y.U.), 24K17104 (S.O.) and 22KJ1990 (R.U.).
This work made use of the JAXA Supercomputer System Generation 3 (JSS3).
 This research has made use of data and/or software provided by the High Energy Astrophysics Science Archive Research Center (HEASARC), which is a service of the Astrophysics Science Division at NASA/GSFC and the High Energy Astrophysics Division of the Smithsonian Astrophysical Observatory. This research has also made use of the NASA/IPAC Extragalactic Database (NED), which is operated by the Jet Propulsion Laboratory, California Institute of Technology, under contract with the
National Aeronautics and Space Administration.
Facilities:  Suzaku (800017010),
NuSTAR (60061228002).
Software: HEAsoft 6.35 (HEASARC 2025), SKIRT \citep{2023A&A...674A.123V},
XSPEC \citep{1996ASPC..101...17A}.

\appendix 
\restartappendixnumbering

\section{IR Photometric Data of NGC 4388}
\subsection{IR Photometric Data}
Table~\ref{c4photo1} summarizes the photometric data of the nucleus in
NGC 4388 used in this work.
The HST data presented in this article were obtained from the Mikulski Archive for Space Telescopes (MAST) at the Space Telescope Science Institute. The specific observations analyzed can be accessed via \dataset[doi:
10.17909/q9xf-tx22] {https://doi.org/10.17909/q9xf-tx22}.

\begin{deluxetable*}{lllllll}
\label{table:SED}
\tablecaption{Nuclear IR Photometry of NGC 4388 \label{c4photo1}}
\tablehead{Telescope&Instrument &Filter &Wavelength &Flux &Reference
\\(1)&(2)&(3)&(4)&(5)&(6)
}
\startdata
    HST    &NICMOS1&F110W          & 1.12$\mu$m &0.06$\pm$0.01&a \\
                   &       &F160W          & 1.60$\mu$m&0.7$\pm$0.1&b \\
            VLT    &SINFONI&K              & 2.25$\mu$m&$<$7.5&c \\
            UKIRT  &IRCAM3 &L              & 3.45$\mu$m&$<$40    &d \\
            GTC	&CanariCam&Si2          & 8.67$\mu$m&74$\pm$11.1&e\\
            IRAS   &       &               &25 $\mu$m&$<$3.57&f\\
                   &       &               &60 $\mu$m&$<$10.27&f\\
                   &       &               &100 $\mu$m&$<$17.15&f\\
            Spitzer&IRS    &               &4.5 $\mu$m&23.5$\pm$3.5&g\\
                   &       &               &5.5 $\mu$m&30.2$\pm$4.5&g\\
                   &       &               &18.0 $\mu$m&788.6$\pm$157.7&g\\
                   &       &               &25.0 $\mu$m&1127.1$\pm$225.4&g\\
                   &       &               &30.0 $\mu$m&1305.7$\pm$261.1&g\\
            SOFIA  &FORCAST&               &31.5 $\mu$m&$<$2040&h\\
\enddata
\tablecomments{
(1) Telescope name.
(2) Instrument name. 
(3) Filter.
(4) Wavelength.
(5) Flux in [mJy].
(6) Reference for IR flux.
}
\tablerefs{
(a).\citealt{2003AJ....126...81A}
(b).\citealt{2001ApJ...547..129Q}
(c).\citealt{2015AA...578A..47B} 
(d).\citealt{1998ApJ...495..196A}
(e).\citealt{2016MNRAS.463.3531G}
(f).\citealt{2003AJ....126.1607S}
(g).\citealt{2019MNRAS.486.4917G}
(h).\citealt{2016MNRAS.462.2618F}
}
\end{deluxetable*}

\subsection{X-ray Spectral Data}

Table \ref{tab-obj} summarizes the X-ray spectral data of the nucleus in NGC 4388 used in this work.

\begin{deluxetable*}{lcccccc}
\tablewidth{\textwidth}
\tablecaption{Observation log \label{tab-obj}}
\tablehead{
Observatory & Observation ID & Start Date & End Date & Exposure (ks) & Ref\\
(1) & (2) & (3) & (4) & (5)  & (6)
}
\startdata
Suzaku & 800017010 & 2005 Dec 24 09:04 & 2005 Dec 27 06:00 & 122 &\citealt{2008PASJ...60S.263S} \\
NuSTAR & 60061228002 & 2013 Dec 27 06:46 & 2013 Dec 27 17:26 & 21 & \citealt{2017ApJ...843...89K}\\
\enddata
\tablecomments{
(1): observatory. (2): observation identification number. (3): start date in units of ymd. (4): end date in units of ymd. 
(5): exposure time in units of kiloseconds, based on good time intervals of XIS 0 for Suzaku and FPMA for NuSTAR.}
\end{deluxetable*}

\section{XCLUMPY(SKIRT)}\label{section:XCLUMPY-SKIRT}

The XCLUMPY \citep{2019ApJ...877...95T} model is calculated using the
MONACO (\citealt{2011ApJ...740..103O, 2016MNRAS.462.2366O})
framework, whereas our IMPACTX model utilizes the SKIRT
framework. We note that there are differences in the atomic database and
physical processes considered between MONACO and SKIRT.
SKIRT adopts
the photoelectric-absorption cross sections by
\citet{1995A&AS..109..125V} and \citet{1996ApJ...465..487V}, whereas
MONACO refers to the xraylib database \citep{2011AcSpB..66..776S}. 
The current SKIRT code does not consider
H$_2$ molecules, whereas MONACO does. Furthermore, the XCLUMPY model
adopts the solar abundance table of \citet{1989GeCoA..53..197A},
whereas our IMPACTX model uses that of \citet{2009LanB...4B..712L}.

To avoid any possible systematic uncertainties caused by these
differences, we have also made a new table model called
``XCLUMPY(SKIRT)'' based on the SKIRT framework, which considers only
the reflection from the torus (i.e., without the polar dusty
gas). This allows us to make direct comparison with the IMPACTX model
created with exactly the same conditions.
Nevertheless, we note that the differences of XCLUMPY(SKIRT) from the
original XCLUMPY model are sufficiently small, and do not affect our
discussion.


\subsection{Fitting Results of NGC 4388}

We reanalyze the Suzaku and NuSTAR spectra using the XCLUMPY(SKIRT) model.
The model is represented as follows in the XSPEC terminology: 

\begin{eqnarray}
\mathrm{model}
&=& \textsf{const1*phabs} \nonumber\\
&*&  \textsf{(const2*zphabs*cabs*zcutoffpl } \nonumber\\
&+& \textsf{const3*zcutoffpl + atable\{xclumpy\_skirt.fits\}} \nonumber\\
&+& \textsf{zgauss + apec1 + apec2)}.
\end{eqnarray}

Here the tables \textsf{xclumpy\_skirt.fits} correspond to the continuum and fluorescent lines in the reflection component from the torus. The other terms are the same as in the fitting model with
IMPACTX described in Section \ref{section:X-fit}. The fitting results
are summarized in Table~\ref{table:XCLUMPY-SKIRT-fit}. We confirm that the parameters are
consistent with those obtained by \citet{2021ApJ...906...84O} using the (original)
XCLUMPY model.


\begin{deluxetable*}{lcCCcc}[htb]\label{table:XCLUMPY-SKIRT-fit}
  \tablecaption{Best fit parameters of the X-ray spectra of NGC 4388 with XCLUMPY(SKIRT)}
 \tablewidth{0pt}
 \tablehead{
 \colhead{Region} & \colhead{No.} & \colhead{Parameter} & \colhead{Best Fit Value} & \colhead{Units}
 }
 \startdata
 Torus&(1) & $\log{N_\mathrm{H}^\mathrm{Equ}}$ & $24.0^{+0.1}_{-0.2}$ & $\mathrm{cm}^{-2}$   \\
        &(2) & $\sigma$ & $19.3^{+9.7}_{-1.3}$ & degree \\
        &(3) & $i$ & $70.0^a$ & degree \\
        &(4) & $Z(\mathrm{Fe})$ & $1.83^{+0.10}_{-0.12}$ & solar \\
        &(5) & $\Gamma_{\mathrm{Suzaku}}$ & $1.48^{+0.04}_{-0.03}$ &  \\
        &(6) & $\Gamma_{\mathrm{NuSTAR}}$ & $1.59^{+0.04}_{-0.02}$ &  \\
        &(7) & $E_{\mathrm{cut}}$ & $370^a$ & keV  \\
        &(8) & $N_X$ & $1.03^{+0.11}_{-0.07}\times{10^{-2}}$ &$\mathrm{photons}~\mathrm{keV}^{-1}~\mathrm{cm}^{-2}~\mathrm{s}^{-1}$ \\
        &(9) & $f_{\mathrm{scat}}$ & $1.67^{+0.10}_{-0.10}\times{10}^{-2}$ &  \\
       Others&(10) & $T_{\mathrm{apec1}}$ & $1.19^{+0.03}_{-0.03}$ & keV \\
        &(11) & $T_{\mathrm{apec2}}$ & $0.31^{+0.02}_{-0.02}$ & keV \\
        &(12) & $N_{\mathrm{apec1}}$& $1.65^{+0.13}_{-0.10}\times{10^{-4}}$ &  \\
        &(13) & $N_{\mathrm{apec2}}$ & $1.32^{+0.14}_{-0.16}\times{10^{-4}}$ &  \\
        &(14) & $C_{\mathrm{TIME}}$ & $0.40^{+0.06}_{-0.06}$ & \\
       &  & $\chi^2/\mathrm{dof}$ & $2023.7/1974$ &  \\
 \enddata
 \tablecomments{
 (1) Hydrogen column density along the equatorial plane. 
 (2) Torus angular width. (3) Inclination angle. (4) Abundance of iron relative to hydrogen. (5)
 Photon index for Suzaku data. (6)
 Photon index for NuSTAR data. (7) Cutoff energy. (8) Normalization of the intrinsic power law component at 1 keV. (9) Scattering fraction.
(10), (11) Temperature of optically thin clouds. (12), (13) Normalization of the optically thin components. (14)Time variability constant.
 } 
 \tablenotetext{a}{The parameter is fixed.}
\end{deluxetable*}


\bibliography{sample631}{}
\bibliographystyle{aasjournal}



\end{document}